# Is Entropy Associated with Time's Arrow?

**Arieh Ben-Naim**

**Department of Physical Chemistry**

**The Hebrew University of Jerusalem**

**Jerusalem, Israel 91904**

**Email address: ariehbennaim@gmail.com**

**Abstract**

We start with reviewing the origin of the idea that entropy and the Second Law are associated with the Arrow of Time. We then introduced a new definition of entropy based on Shannon's Measure of Information (SMI). The SMI may be defined on any probability distribution; and therefore it is a very general concept. On the other hand entropy is defined on a very special set of probability distributions. More specifically the entropy of a thermodynamic system is related the probability distribution of locations and velocities (or momenta) of all the particles, which maximized the Shannon Measure of Information. As such, entropy is not a function of time. We also show that the H-function, as defined by Boltzmann is an SMI but not entropy. Therefore, while the H-function, as an SMI may change with time, Entropy, as a limit of the SMI does not change with time.

Keywords: Arrow of Time, Entropy, Shannon's measure of information, Second Law of Thermodynamics, H-Theorem, H-function.

# 1. Introduction

Open any book which deals with a "theory of time," "time's beginning," and "time's ending," and you are likely to find the association of entropy and the Second Law of Thermodynamics with time [1-17]. The origin of this association of "Time's Arrow" with entropy can be traced to Clausius' famous statement of the Second Law [18]:

"*The entropy of the universe always increases.*"

The statement "*entropy always increases,*" always means that "*entropy always increases with time*." In this article, we show that statements like "*entropy always increases*" are meaningless; entropy, in itself does not have a numerical value, therefore one cannot claim that it increases or decreases. Entropy changes are meaningful only for well-defined thermodynamic systems for which the entropy is defined; once we specify the system, then its entropy is determined, and does not change with time.

Eddington [19] is credited for the explicit association of "The law that entropy always increases" with "Time's Arrow," which expresses this "one-way property of time." Quotations from Eddington feature in most popular science books, as well as in some textbooks on thermodynamics. Here are the two relevant quotations from Eddington's (1928) book [19], "The Nature of the Physical World." The first concerns the role of entropy and the Second Law, and the second, introduces the idea of "time's arrow."

  1.  *"The law that entropy always increases, holds, I think, the supreme position among the laws of Nature".*

  2.  *"Let us draw an arrow arbitrarily. If as we follow the arrow we find more and more of the random element in the state of the world, then the arrow is pointing towards the future; if the random element decreases the arrow points towards the past.*

*This follows at once if our fundamental contention is admitted that the introduction of randomness is the only thing which cannot be undone. I shall use the phrase 'time's arrow' to express this one-way property of time which has no analogue in space".*

In the first quotation Eddington reiterates the unfounded idea that "entropy always increases." Although it is not explicitly stated, the second quotation alludes to the connection between the Second Law and the Arrow of Time. This is clear from the (erroneous) association of the "random element in the state of the world" with the "arrow pointing towards the future."

In my view it is far from clear that an Arrow of Time exists [11, 20, 21]. It is also clear that entropy is not associated with randomness, and it is far from clear that entropy always increases [21-24].

Here is another typical quotation from Rifkin's book [16], associating *entropy* with the *Arrow of Time*.

  *"...the second law. It is the irreversible process of dissipation of energy in the world. What does it mean to say, 'The world is running out of time'? Simply this: we experience the passage of time by the*

*succession of one event after another. And every time an event occurs anywhere in this world energy is expended and the overall entropy is increased. To say the world is running out of time then, to say the world is running out of usable energy. In the words of Sir Arthur Eddington, 'Entropy is time's arrow."*

Such statements were driven to the extremes of absurdity by using the *quality sign* "=" to express the identity of time's arrow and entropy. In "The Demon and the Quantum," Scully writes [13]:

"*The statistical time concept that entropy = time's arrow has deep and fascinating implications. It therefore behooves us to try to understand entropy ever more deeply. Entropy not only explains the arrow of time, it also explains its existence; it is time."*

Most writers on entropy and the Second Law, adhering to Clausius' statement of the Second Law claim that since the entropy of the universe always increases, it follows that the entropy of the universe must have been low at, or near, the Big Bang. This is known as the Past Hypothesis [1,5]. Some authors [5] even go a step further, and use the Past Hypothesis to *explain* everything that happens, including our ability to remember the past but not the future.

In this article, we focus on the question posed in the article's title. We shall discuss the more specific concepts such as the "entropy of the universe," or the "entropy of a living system" in a future article.

In the next section, we introduce very briefly three definitions of entropy and the corresponding formulations of the Second Law of thermodynamics. The first, due to Clausius, the second, due to Boltzmann, and the third, based on Shannon's measure of information (SMI). We shall see that one of the main advantages of using the third definition of entropy is that it shows clearly why entropy *is not* a function of time, and cannot be identified with the "Arrow of Time."

The main result of this definition of entropy is that entropy is proportional to the *maximum* of the SMI; maximum over all the probability distributions of locations and momenta of the particles (see Appendix A).

$$S = K \, \text{Max}_{all \, distributions \, f} \, \text{SMI} \, [f(R,p)] \qquad (1.1)$$

Thus, instead of saying that the entropy changes with time and reaches a maximum at equilibrium, we shall show that the correct statement is that SMI (associated with the distributions of locations and momenta) changes with time, and reaches a maximum at equilibrium. The *value* of the entropy is identified with the limit of the SMI at $t \to \infty$, or equivalently when the system reaches equilibrium. In other words, we identify the maximum in eq. (1.1), over all possible distributions, with the time limit

$$S = K \lim_{t \to \infty} \text{SMI}[f(R,p,t)] \qquad (1.2)$$

From this result it is clear that entropy is not a function of time!

This article concludes with three recommendations:

1. It is essential to make a clear-cut distinction between SMI and entropy. Failure to make such a distinction has caused a great confusion in both thermodynamics and Information theory.

2. To reserve and use the concept of entropy only to macroscopic systems at *equilibrium*, and to use the concept of SMI for all other systems; small or large number of particles, near or far from equilibrium.

3. The identification of the maximum in equation (1.1) with the limit in (1.2) is tentative. It is based on the (unproven) existence of the time limit in (1.2). The *state* of the system (as specified by the probability distribution of the locations and momenta of all particles) can fluctuate even when the system reaches equilibrium. The entropy of the system, defined in eq. (1.1), on the other hand, does not fluctuate for a well-defined thermodynamic system.

## 2. The three definitions of entropy

In this section, we shall briefly discuss three definitions of entropy, a more detailed discussion may be found in references [25-28]

### 2.1 Clausius' definition

For our purpose, in this article it is clear that Clausius' definition, together with the third law determines the entropy of a system at *equilibrium*. Clausius' definition cannot be used to extend the definition of entropy to non-equilibrium states, Appendix B. This observation is important. As we shall see, any other definition will be accepted as a valid definition of entropy only when calculations based on that definition agrees with the calculations based on Clausius' definition.

### 2.2 Boltzmann's definition

There are essentially two "definitions" of the Boltzmann entropy [29-38]. One is the famous equation

$$S_B = k_B \log W \ ,$$ (2.1)

where $k_B$ is the Boltzmann constant, and $W$ is the number of accessible microstates of a system having a fixed energy, volume, and number of particles. This equation is based on the assumption that all the microstates of the system are equally probable. From this definition of entropy Gibbs derived an expression of the entropy for systems described by other macroscopic variables such as: $(T, V, N), (T, P, N)$ or $(T, V, \mu)$. All these entropies have the same *formal* form [38]

$$S_G = -k_B \sum_i p_i \log p_i$$ (2.2)

where $p_i$ are the equilibrium probabilities of finding the system with fixed energy (in a $T, V, N$ ensemble), fixed energy and volume (in a $T, P, N$ ensemble), or fixed energy and number of particles (in a $T, V, \mu$ ensemble, where $\mu$ is the chemical potential).

It is clear that both $S_B$ in (2.1) and $S_G$ in (2.2) are time independent. Also, for any system for which one can calculate the changes in entropy from either (2.1) or (2.2), [for instance, for an expansion of an ideal gas, or mixing of ideal gases], one finds agreement between these calculations and the calculation based on Clausius' definition. We emphasize again that all these calculations of entropy changes pertain to equilibrium systems, see also section 3 below.

The second, so-called Boltzmann entropy is based on Boltzmann's definition of the H-function [37]:

$$H(t) = \int f(v,t) \log f(v,t) dv \qquad (2.3)$$

This function is obviously a function of time, and is defined for any distribution of velocities $f(v,t)$. Boltzmann identified the function $-H(t)$ with entropy. This identification is prevalent in the literature even today [17, 36, 37].

In the following, we shall see that this identification is not valid, mainly because it cannot be compared with calculations based on Clausius' definition. We shall return to this misidentification in section 4 below.

### 2.3 Definition of entropy based on Shannon's measure of information (SMI)

In this section, we present very briefly the third definition of entropy based on the SMI. More details are provided in Appendix A and in references [20, 22, 24, 27, 28].

We start with the SMI, as defined by Shannon [39]. The SMI is defined for any given probability distribution $p_1, p_2, \ldots, p_N$ by

$$H = -K \sum p_i \log p_i \qquad (2.4)$$

where $K$ is a positive constant and the logarithm was originally taken to the base 2. Here, we use the natural logarithm and include into $K$ the conversion factor between any two bases of the logarithm.

Clearly, the SMI defined in eq. (2.4) is a very general quantity. It is defined for any probability distribution; it can be the probabilities of Head and Tail for tossing a coin, or the six outcomes of a die. It is unfortunate that because of the formal resemblance of (2.4) to Gibbs entropy, the SMI is also referred to as entropy. This has caused a great confusion in both information theory and thermodynamics. This confusion was already recognized by Jaynes who initially adopted the term "entropy" for the SMI, but later realized the potential confusion it could create [27, 28 , 39-43].

In order to obtain the thermodynamic entropy $S$ from the SMI we have to start with the SMI and proceed in two steps: First, apply the SMI to the probability distribution of *locations* and *momenta* of a system of many particles. Second, calculate the maximum of the SMI over all possible such distributions. For more details, see Appendix A. We note here that in this derivation we use the *continuous* analogue of the SMI written as:

$$\text{SMI}(locations\ and\ velocities) = -K \int f(\boldsymbol{R}, \boldsymbol{v}, t) \log f(\boldsymbol{R}, \boldsymbol{v}, t) d\boldsymbol{R} d\boldsymbol{v} \qquad (2.5)$$

However, in actual applications for thermodynamics we always use the discrete definition of the SMI as shown in (2.4). See Appendix A.

The procedure of calculating the distribution that maximizes the SMI in (2.5) is known as the MaxEnt (maxium entropy) principle. We shall refer to this procedure as the MaxSMI, and not as MaxEnt [44].

For a system of non-interacting particles the distribution of locations is uniform and that of velocities (or momenta) is Maxwell Boltzmann [20, 24]. This was originally shown by Shannon himself and some further details are provided in Appendix A and in references [20, 24].

After finding the distribution that maximizes SMI in (2.5), denoted by $f^*(\boldsymbol{R}, \boldsymbol{v})$, we can calculate the maximum SMI for that particular distribution, i.e.

$$\text{MaxSMI} = -K \int f^*(\boldsymbol{R}, \boldsymbol{v}) \log f^*(\boldsymbol{R}, \boldsymbol{v}) d\boldsymbol{R} d\boldsymbol{v} \qquad (2.6)$$

Once we calculate the MaxSMI for an ideal gas, we find that the value of the MaxSMI is the same as the entropy of an ideal gas as calculated by Sackur and Tetrode [20, 24, 45, 46] , which is the entropy of an ideal gas at a specified $E$, $V$ and $N$ at equilibrium. Therefore, we can *define* the entropy of an ideal gas, up to a multiplicative factor $K$, as the MaxSMI as defined in (2.5). Note that unlike the distribution $f(\boldsymbol{R}, \boldsymbol{v}, t)$ in eq. (2.5) the distribution which maximizes the SMI, denote $f^*(\boldsymbol{R}, \boldsymbol{v})$ is not a function of time.

Furthermore, since we know that the locational distribution of an ideal gas at equilibrium is the uniform distribution, and that the velocity distribution of an ideal gas at equilibrium is the Maxwell-Boltzmann distribution, we can identify the distribution $f^*$ in (2.6) which *maximizes* the SMI, with the *equilibrium* distribution $f^{eq}(\boldsymbol{R}, \boldsymbol{v})$, and instead of (2.6) we write

$$\text{MaxSMI} = -\int f^{eq}(\boldsymbol{R}, \boldsymbol{v}) \log f^{eq}(\boldsymbol{R}, \boldsymbol{v}) d\boldsymbol{R} d\boldsymbol{v} \qquad (2.7)$$

Clearly, this MaxSMI *is not* a *function* of *time*. Thus, this definition of the entropy of an ideal gas is equivalent to the definition of Boltzmann's entropy, as well as Clausius' entropy. Equivalent, in the sense that calculations of entropy changes between two *equilibrium states* of an ideal gas, based on this definition agree with the result based on Clausius' entropy.

One can also extend the definition of entropy based on the MaxSMI, to a system of interacting particles [20, 24]. In which case, the interactions between the particles produce *correlations*, which in turn can be cast in the form of *mutual* information between the particles [20, 24].

Thus, the procedure of MaxSMI also provides a definition of entropy for systems consisting of interacting particles at equilibrium.

## 3. Expansion of an ideal gas

In this section, we use the simplest example of a spontaneous process for which we can calculate the change in entropy and answer a few questions relevant to the Second Law. All the conclusions reached in this section are valid to any other thermodynamic system at equilibrium.

Consider the simple example of expansion of $N$ particles from volume $V$ to $2V$, Figure 1. We also assume that the particles are *simple*, i.e. they have no internal degrees of freedom.

For any $N$, right after removing the partition we follow the evolution of the system with time. We observed that the particles which were initially confined to one compartment of volume $V$, can access the larger volume $2V$. We can ask the following questions:

1. Why does the gas expand to occupy the larger volume?
2. Does the number of particles in the left compartment change *monotonically* with time?
3. Does the number of particles in the left compartment reach a constant value at equilibrium?
4. How fast does the system reach the equilibrium state?
5. How does the SMI of the system change with time?
6. How does the entropy of the system change as a result of this process?

Clearly, the answers to all these questions depend on $N$. In the following we provide the answers to these questions for the particular process of expansion of an ideal gas. Similar answers are valid for a process of mixing two ideal gases as shown in Figure 2, [20, 24, 47, 48].

1. The answer to the first question is probabilistic [27, 28]. The reason the particles will occupy the larger volume $2V$ rather than $V$ is that the probability of the states where there are about $N/2$ in each compartment is larger, than the probability of the state where all the particles are in one compartment. This is true for any $N$. However, when $N$ is very small, there is a relatively large probability that the particles will be found in one compartment. For these cases we cannot claim that the process is *irreversible*, in the sense that it will never go back to the initial state. For large $N$, even of the order of $10^6$, the probability of returning to the initial state becomes so small, that it is practically zero. However, there is always a *finite* probability that the system will visit the initial state. For $N$ or the order of $10^{23}$, the probability of visiting the initial state is so small (but still non-zero) that we can safely say that the system will *never* return to the initial state. *Never,* here means in the sense of *billions* of *ages* of the universe. This is the source of the term "irreversibility" assigned to this process or to the mixing of ideal gases, Figure 2.

For a system of non-interacting particles, we can calculate the probability of finding any distribution of the $N$ particles in the two compartments: $\left(N_1(t),\ N - N_1(t)\right)$, where $N_1(t)$ is the number of particles in the left compartment at time $t$ after the removal of the partition. The probability of finding such a distribution is

$$\Pr(t) = \Pr[N_1(t),\ N - N_1(t)] \ = \left(\tfrac{1}{2}\right)^N \binom{N}{N_1(t)} \qquad (3.1)$$

Clearly, since the probability of finding any specific particle in either the left or the right compartment is $\frac{1}{2}$, the total number of configuration is $2^N$ and the number of configurations for which there are $N_1(t)$ particles in one compartment is $\begin{pmatrix} N \\ N_1(t) \end{pmatrix}$. We define the probabilities: $p_1(t) = \frac{N_1(t)}{N}$, and $p_2(t) = \frac{N-N_1(t)}{N}$, and rewrite equation (3.1) as

$$\Pr[p_1(t), p_2(t)] = \left(\frac{1}{2}\right)^N \begin{pmatrix} N \\ N\,p_1(t) \end{pmatrix} = \left(\frac{1}{2}\right)^N \frac{N!}{[Np_1(t)]![Np_2(t)]!} \qquad (3.2)$$

In this form we expressed the probability Pr as a function of the probability distribution; $[p_1(t), p_2(t)]$. This probability has a maximum when $p_1(t) = p_2(t) = 1/2$. Clearly, the answer to the first question is probabilistic. The probability of having about equal number of particles in the two compartments is always larger than the probability of finding all particles in one compartment. More details on this in references [20, 24].

2. The answer to the second question is, in general, No; the number of particles in L does not change monotonically from $N$ to $N/2$ (or from zero to $N/2$ if we start with all particles in the right compartments). Simulations show that for large values of $N$ the number $n$ changes *nearly* monotonically towards $N/2$. The larger the $N$, the more nearly monotonic is the change of $N_1(t)$. (For simulated results, see reference [47] and arienbennaim.com, books, Entropy Demystified, simulated games). For $N$ on the order of $10^6$ or more, one will see nearly perfect, smooth, monotonic change in the number of particles in L.

3. The answer to the third question depends on how one defines the equilibrium state of the system. If we define the equilibrium state when the value of $N_1(t)$ is equal to $N/2$, then when $N_1(t)$ reaches $N/2$ it *will not* stay there "forever." There will always be fluctuations about the value of $N/2$. However, one can define the equilibrium state as the state for which $N_1(t)$ is in the *neighborhood* of $N/2$. In such a definition, we will find that once $N_1(t)$ reaches this neighborhood; it will stay there for a longer time than in any other state [20, 24]. For $N$ of the order of $10^6$ or more, the system will stay in this neighborhood "forever." Forever, means here many ages of the universe.

4. The answer to fourth question depends on the temperature and on the size of the aperture we open between the two compartments. In the experiment of Figure 1 we remove the partition between the two compartments. However, we could do the same experiment by opening a small window. In such an experiment, the speed of reaching the equilibrium state would depend also on the size of the aperture of the window. In any case thermodynamics does not say anything about the *speed* of attaining equilibrium.

5. The answer to the fifth question is most relevant to the probabilistic formulation of the Second Law [27, 28]. For each distribution of particles $(N_1(t), N_2(t) = N - N_1(t))$ we can define a probability distribution: $(p_1(t), p_2(t))$, and the corresponding SMI. As the system evolves from the initial to the final state, $N_1(t)$ will change with time, hence also $p_1(t)$ will change with time, hence also the SMI will change with time. (For simulations, see reference [47]). The relationship between the SMI and the probability may be calculated as follows:

Using the Stirling approximation in the form $\ln n! \approx n\ln n - n$ for each factorial on the right hand side of (3.2), we obtain

$$\Pr(t) = \Pr[p_1(t), p_2(t)] = \left(\frac{1}{2}\right)^N 2^{N \times \mathrm{SMI}[p_1(t), p_2(t)]} \qquad (3.3)$$

Thus, at any time $t$ we have a distribution of particles in the two compartments; $N_1(t), N_2(t)$. Each distribution of particles defines a probability distribution $[p_1(t), p_2(t)]$. On this probability distribution one can define a probability function, $\Pr(t) = \Pr[p_1(t), p_2(t)]$, as well as a SMI($t$), which are time dependent. The relationship between the probability Pr and the SMI is shown in eq. (3.3).

A more general relationship exists for a system of any number of compartments and of any number of particles. It is easy to show that Pr has a maximum when $p_1(t) = p_2(t) = \frac{1}{2}$, i.e. when there are N/2 particles in each compartment. At this distribution, both the probability $\Pr(t)$, and the SMI($t$) attains maximum values. Specifically,

$$\mathrm{MaxSMI} = -\frac{1}{2}\log_2 \frac{1}{2} - \frac{1}{2}\log_2 \frac{1}{2} = 1 \qquad (3.4)$$

Thus, we have for the ratio of the probabilities at the two distributions $\left[\frac{1}{2}, \frac{1}{2}\right]$, and [1,0]:

$$\frac{\Pr\left[\frac{1}{2}, \frac{1}{2}\right]}{\Pr[1,0]} \approx 2^{N\left(\mathrm{SMI}\left[\frac{1}{2}, \frac{1}{2}\right] - \mathrm{SMI}[1,0]\right)} = 2^N = e^{N\ln 2} \qquad (3.5)$$

The conclusion from equation (3.3) is that both the probability Pr and the SMI change with time $t$. However, at the final equilibrium state the ratio of the probabilities in eq. (3.5) is related to the difference in the SMI, which, in turn is related to the difference in the entropies of the system between the final and the initial states:

$$S(final) - S(initial) = Nk_B \ln 2 \qquad (3.6)$$

This quantity, up to the Boltzmann constant $k_B$, is equal to the change in the SMI. Note that the entropy change in (3.6) may be calculated from any of the three definitions of the entropy. Here, we got it from the change in the SMI in the process shown in Figure 1.

For small $N$, the SMI will start from zero (all particles being in one compartment) and will fluctuate between zero to $N$ bits. When $N$ is very large, say $10^6$ or more the value of SMI will change nearly monotonically from zero to $N$ bits. There will always be some fluctuations in the value of SMI, but these fluctuations will be smaller the larger $N$. Once the system reaches the equilibrium state it will stay there *forever*. Note carefully that the SMI is defined here on the probability distribution $\left(p_1(t), p_2(t)\right)$. For the initial distribution $(1,0)$ the SMI is zero. The SMI defined on the distribution of locations and momenta is not zero.

6. The answer to the last question is the simplest, yet it is the most misconstrued one. It is the simplest because entropy is a *state* function; it is defined for a well-defined state of a macroscopic system. For the expansion process, the *macrostate* of the system is defined initially by $(E, V, N)$. The corresponding value of the entropy is $S(E, V, N)$. The final macrostate is characterized by $(E, 2V, N)$, and the corresponding value of the entropy is $S(E, 2V, N)$. The change in entropy is given by equation (3.6). In between the two macrostates $(E, V, N)$ and $(E, 2V, N)$, the macrostate of the system is not well-defined. A few, intermediate states are shown in Figure 3. While $E$ and $N$ are the same as in the initial state, the "volume" during the expansion process of the gas is not well-defined. It becomes well-defined only when the system reaches an equilibrium state. Therefore, since the volume of the system is not well-defined while the gas expands, also the entropy is not well-defined. We can say that the entropy changes abruptly from $S(E, V, N)$ to $S(E, 2V, N)$, and that this change occurred at the moment the system reaches the final equilibrium state. At all intermediate states the entropy of the system is not defined. This is schematically shown in Figure 4a.

One can also adopt the point of view that when we remove the partition between the two compartments, the volume of the gas changes abruptly from $V$ to $2V$, although the gas is initially still in one compartment, the total volume $2V$ is *accessible* to all particles. If we adopt this view, then at the moment we removed the partition, the volume changes from $V$ to $2V$, and the corresponding change in entropy is $S(E, 2V, N) - S(E, V, N)$. This change occurs abruptly at the moment we remove the partition, see Figure 4b. Personally, I prefer the first point of view. Initially, the system has the value $S(E, V, N)$ before the removal of the partition, and it reaches the value of $S(E, 2V, N)$ after the systems reached the new, final equilibrium state. In all the intermediate states the entropy is not defined. Note however, that the SMI is defined for any intermediate states between the initial and the final states. However, the entropy is the maximum value of the SMI (multiplied by the Boltzmann constant and change of the base of the logarithm), reached at the new equilibrium state.

It should be noted however that we could devise another expansion (referred to as quasi-static) process by moving gradually the partition between the two compartments. In this process the system proceeds through a series of equilibrium states, and therefore the entropy is well-defined at each of the points along the path leading from $(E, V, N)$ to $(E, 2V, N)$. In this process, the entropy of the gas will gradually change from $S(E, V, N)$ to $S(E, 2V, N)$, Figure 5. The length of time it takes to proceed from the initial to the final state depends on how fast, or how slow we decide carry out the process.

Finally, we note that in this particular process the average kinetic energy of the system does not change. The change in the SMI as well as in the entropy is due to the change in the *locational* distribution of the particles. In the next section, we discuss the case of evolution of the velocity distribution. This case is discussed within the so-called Boltzmann's H-Theorem.

## 4. Boltzmann's H-theorem; the criticism and its resolution

Before we discuss Boltzmann's H-theorem, we summarize here the most important conclusion regarding the SMI.

In Section 3, we saw that the entropy is obtained from the SMI in four steps. We also saw that the entropy of a thermodynamic system is related to the *maximum* value of the SMI defined on the distribution of locations and velocities of all particles in the system:

$$S = K \text{ MaxSMI}(locations\ and\ velocities) \qquad (4.1)$$

where $K$ is a constant ($K = k_B \ln 2$).

We know that every system tends to an equilibrium state at very long time, therefore we identify the MaxSMI as the time limit of the SMI, i.e.

$$S = K \lim_{t\to\infty} \text{SMI}(locations\ and\ velocities) \qquad (4.2)$$

The derivation of the entropy from the SMI is a very remarkable result. But what is more important is that this derivation reveals at the same time the relationship between entropy and SMI on one hand, and the fundamental difference between the two concepts, on the other hand.

Besides the fact that the SMI is a far more general concept than entropy, we found that even when the two concepts apply to the distribution of locations and velocities, they are different. The SMI can evolve with time and reaches a limiting value (for large systems) at $t \to \infty$.

The entropy is proportional to the maximum value of the SMI obtained at equilibrium. As such entropy is not, and cannot be a function of time. Thus, the "well-known" mystery about the "entropy always increase with time," disappears. With this removal of the mystery, we also arrive at the resolution of the "paradoxes" associated with the Boltzmann H-theorem.

There have been many attempts to *derive* the Second Law of Thermodynamics from the dynamics of the particles. Mackey[17] devote a whole book: "Time's Arrow, the Origins of Thermodynamic Behavior" to this question. In fact, the first attempt to derive an equation for the "entropy" of a system which changes with time and reaches a maximum at equilibrium was done by Boltzmann in his famous H-theorem.

In 1877, Boltzmann defined a function $H(t)$ [29, 30]

$$H(t) = \int f(v,t) \log[f(v,t)]\,dv \qquad (4.3)$$

Boltzmann proved a remarkable theorem known as Boltzmann's H-theorem. Boltzmann made the following assumptions:

1. Ignoring the molecular structure of the walls (ideal. perfect smooth walls).

2. Spatial homogenous system or uniform locational distribution.

3. Assuming binary collisions, conserving momentum and kinetic energy

4. No correlations between location and velocity (assumption of molecular chaos).

Then, Boltzmann proved that:

$$\frac{dH(t)}{dT} \leq 0 \qquad\qquad (4.4)$$

and at equilibrium, i.e., $t \to \infty$:

$$\frac{dH(t)}{dT} = 0 \qquad\qquad (4.5)$$

Boltzmann believed that the behavior of the function $-H(t)$ is the same as that of the entropy, i.e., the entropy always increases with time, and at equilibrium, it reaches a maximum. At this time, the entropy does not change with time. This theorem drew a great amount of criticism, the most well-known are:

*I. The "Reversal Paradox" States:*

"The $H$-theorem singles out a preferred direction of time. This is inconsistent with the time reversal invariance of the equations of motion".

This is not a paradox because the statement that $H(t)$ always changes in one direction is false.

*II. The "Recurrence Paradox," Based on Poincare's Theorem States:*

"After sufficiently long time, an isolated system with fixed *E, V, N*, will return to arbitrary small neighborhood of almost any given initial state."

If we assume that $dH/dT < 0$ at all *t,* then obviously $H$ cannot be periodic function of time.

Both paradoxes have been with us ever since. Furthermore, most popular science books identify the Second Law, or the behavior of the entropy with the so-called *arrow* of *time*. Both paradoxes seem to arise from the conflict between the *reversibility* of the equations of motion on one hand, and the apparent *irreversibility* of the Second Law, namely that the H-function decreases monotonically with time. Boltzmann rejected the criticism by claiming that H does not always decrease with time, but only with high probability. The irreversibility of the Second Law is not absolute, but also highly improbable. The answer to the recurrence paradox follows from the same argument. Indeed, the system can return to the initial state. However, the recurrence time is so large that this is never observed, not in our lifetime, not even in the life time of the universe.

Notwithstanding Boltzmann's correct answers to his critics, Boltzmann and his critics made an enduring mistake in the H-theorem, a lingering mistake that has hounded us ever since. This is the very identification of the function $-H(t)$ with the behavior of the entropy. This error has been propagated in the literatures until today.

It is clear, from the very definition of the function $H(t)$, that $-H(t)$ is a SMI. And if one identifies the SMI with entropy, then we go back to Boltzmann's identification of the function $-H(t)$ with entropy.

Fortunately, thanks to the recent derivation of the *entropy function*, i.e. the function $S(E, V, N)$, or the Sackur-Tetrode equation for the entropy based on the SMI, it becomes crystal clear that the SMI is not entropy! The entropy is obtained from the SMI when applied to the distribution of locations and momenta, then take the limit $t \to \infty$, and only in this limit we get entropy function which has no traces of time dependence.

Translating our findings in Section 3 to the H-theorem, we can conclude that $-H(t)$ is SMI based on the velocity distribution. Clearly, one cannot identify $-H(t)$ with entropy. To obtain the entropy one must first define the $-H(t)$ function based on the distribution of both the locations and momentum, i.e.

$$-H(t) = -\int f(\boldsymbol{R}, \boldsymbol{p}, t) \log f(\boldsymbol{R}, \boldsymbol{p}, t) d\boldsymbol{R} d\boldsymbol{p} \qquad (4.6)$$

This is a proper SMI. This may be defined for a system at equilibrium, or very far from equilibrium. To obtain the entropy one must take the limit $t \to \infty$, i.e., the limit $-H(t)$ at equilibrium, i.e.:

$$\lim_{t \to \infty} [-H(t)] = \text{MaxSMI} \; (at \; equilibrium) \qquad (4.7)$$

At this limit we obtain the entropy (up to a multiplicative constant), which is clearly not a function of time.

Thus, once it is understood that the function $-H(t)$ is an SMI and not entropy, it becomes clear that the criticism of Boltzmann's H-Theorem were addressed to the evolution of the SMI and not to the entropy. At the same time, Boltzmann was right in defending his H-theorem when viewed as a theorem on the evolution of SMI, but he was wrong in his interpretation of the quantity $-H(t)$ as entropy.

## 5. Conclusion

Boltzmann's contribution to understanding entropy and the Second Law is undeniable and unshakable. However, Boltzmann also contributed to the lingering misinterpretation of entropy and the Second Law. The first misinterpretation is associating entropy with disorder.

Boltzmann was probably the first to associate entropy with disorder. Here are some quotations [32].

*"… the initial state of the system…must be distinguished by a special property (ordered or improbable)…"*

*"…this system takes in the course of time states…which one calls disordered."*

*"Since by far most of the states of the system are disordered, one calls the latter the probable states."*

*"… the system…when left to itself, it rapidly proceeds to the disordered, most probable state."*

Boltzmann never "equated" entropy with disorder [49], as many others did. However, the "disordered" interpretation of entropy has been with us until today. Criticism of this interpretation has been published earlier [20, 24, 27].

The second Boltzmann's failure is associated with his H-function and H-theorem. As we have noted in section 4, the H-theorem drew a lot of criticisms which were successfully repelled by Boltzmann himself. The main argument of defending his H-theorem is based on the idea that the Second Law is basically a law of probability. But Boltzmann, as many others failed to pinpoint the *subject* on which probability operated: In other words, the failure to distinguish between the two statements:

1. The state of the thermodynamics changes from lower probability to a higher probability.

2. The entropy of the system changes with high probability towards a maximum.

Whilst the first statement is correct, and in fact applies to any thermodynamic system (isolated, isothermal, isothermal isobaric, etc.), the second is incorrect. First, the entropy formulation of the Second Law only applies to isolated systems, and second, the correct formulation is that the SMI of an isolated system tends to a maximum at equilibrium, and at equilibrium the value of the MaxSMI is proportional to the entropy of the system.

How do we know this? We know this, thanks to Shannon's measure of information. It is ironic that Shannon named his quantity entropy [50], and by doing so he contributed to great confusion in both thermodynamics and information theory. However, from two theorems proved by Shannon, it follows that the entropy (the thermodynamic entropy) of the system is attained at a distribution (of locations and velocities) which *maximizes* the SMI. We also know that the distribution of locations of ideal gases is uniform, and the distribution of velocities is the MB distribution. Therefore, we can identify the maximal value of the SMI with the value of the SMI at equilibrium. Since there is only one distribution that maximizes the SMI, it follows that the entropy of a thermodynamic system has a unique value. It does not change with time, it does not reach a maximum value with time, and it does not fluctuate at equilibrium (neither with low or high probability).

In many popular science books one can find plots showing how entropy changes with time [5,8,12]. Most of the time fluctuations are small, but once in many billions of years it might have a big fluctuation.

Another misconception associated with time-dependence of entropy is associated with the so-called Past Hypothesis [1,5], namely that since entropy of the universe always increases, one can extrapolate back and conclude that the entropy of the universe at the Big Bang must have been very low. The second unwarranted "prediction" may be referred to as the Future Hypothesis which basically states that the universe is doomed to "thermal death." These two hypotheses are unwarranted. They were criticized in reference [21], and will be discussed further in a future article.

We conclude this article with two recommendations:

1. It is essential to make a clear-cut distinction between SMI and Entropy. Failure to make such a distinction has caused great confusion in both thermodynamics and Information Theory.

2. To reserve and use the concept of entropy only to macroscopic systems at *equilibrium*, and to use the concept of SMI for all other systems; small or large number of particles, near or far from equilibrium.

## Appendix A.  Definition of Entropy bases on Shannon's Measure of Information

In this Appendix we derive the entropy function for an ideal gas. We start with SMI which is definable to any probability distribution. We apply the SMI to two specific molecular distributions; the locational and the momentum distribution of all particles. Next, we calculate the distribution which maximizes the SMI. We refer to this distribution as the *equilibrium* distribution. Finally, we apply two corrections to the SMI, one due to Heisenberg uncertainty principle, the second due to the indistinguishability of the particles. The resulting SMI is, up to a multiplicative constant equal to the entropy of the gas, as calculated by Sackur and Tetrode based on Boltzmann definition of entropy [11,12].

In previous publication [2,13], we discussed several advantages to the SMI-based definition of entropy. For our purpose in this article the most important aspect of this definition is the following:

The entropy is *defined* as the maximum value of the SMI. As such, it is not a function of time.

### A.1. The Locational SMI of a Particle in a 1D Box of Length L

Suppose we have a particle confined to a one-dimensional (1D) "box" of length L. Since there are infinite points in which the particle can be within the interval (0, L). The corresponding locational SMI must be infinity. However we can defined, as Shannon did, the following quantity by analogy with the discrete case:

$$H[f] = -\int f(x) \log f(x) dx \qquad (A.1)$$

This quantity might either converge or diverge, but in any case, in practice we shall use only differences of this quantity. It is easy to calculate the density which maximizes the locational SMI, $H(f)$ in (A.1) which is [20,24]:

$$f_{eq}(x) = \frac{1}{L} \qquad (A.2)$$

The use of the subscript eq (for equilibrium) will be cleared later, and the corresponding SMI calculated by (A.1) is:

$$H_{max}(\text{locations in } 1D) = \log L \qquad (A.3)$$

We acknowledge that the location X of the particle cannot be determined with absolute accuracy, i.e. there exists a small interval, $h_x$ within which we do not care where the particle is. Therefore, we must correct equation (A.3) by subtracting $\log h_x$. Thus, we write instead of (A.3):

$$H(location \ in \ 1D) = \log L - \log h_x \qquad (A.4)$$

We recognize that in (A.4) we effectively defined SMI for a finite number of intervals $n = L/h$. Note that when $h_x \to 0$, $H$ in (A.4) diverges to infinity. Here, we do not take the mathematical limit, but we stop at $h_x$ small enough but not zero. Note also that in writing (A.4) we do not have to specify the units of length, as long as we use the same units for L and $h_x$.

## A.2. The Velocity SMI of a Particle in 1D "Box" of Length L

Next, we calculate the probability distribution that maximizes the continuous SMI, subject to two conditions:

$$\int_{-\infty}^{\infty} f(x)dx = 1 \qquad (A.5)$$

$$\int_{-\infty}^{\infty} x^2 f(x)dx = \sigma^2 = constant \qquad (A.6)$$

The result is the Normal distribution [20, 24]:

$$f_{eq}(x) = \frac{\exp[-x^2/\sigma^2]}{\sqrt{2\pi\sigma^2}} \qquad (A.7)$$

The subscript eq. for equilibrium will be clear later. Applying this result to a classical particle having average kinetic energy $\frac{m<v_x^2>}{2}$, and identifying the standard deviation $\sigma^2$ with the temperature of the system:

$$\sigma^2 = \frac{k_B T}{m} \qquad (A.8)$$

We get the equilibrium velocity distribution of one particle in 1D system:

$$f_{eq}(v_x) = \sqrt{\frac{m}{2mk_B T}} \exp\left[\frac{-mv_x^2}{2k_B T}\right] \qquad (A.9)$$

where $k_B$ is the Boltzmann constant, $m$ is the mass of the particle, and $T$ the absolute temperature. The value of the continuous SMI for this probability density is:

$$H_{max}(\text{velocity in } 1D) = \frac{1}{2}\log(2\pi e k_B T/m) \qquad (A.10)$$

Similarly, we can write the momentum distribution in 1D, by transforming from $v_x \to p_x = mv_x$, to get:

$$f_{eq}(p_x) = \frac{1}{\sqrt{2\pi m k_B T}} \exp\left[\frac{-p_x^2}{2mk_B T}\right] \qquad (A.11)$$

and the corresponding maximal SMI:

$$H_{max}(\text{momentum in } 1D) = \frac{1}{2}\log(2\pi e m k_B T) \qquad (A.12)$$

As we have noted in connection with the locational SMI, the quantities (A.11) and (A.12) were calculated using the definition of the *continuous* SMI. Again, recognizing the fact that there is a limit to the accuracy within which we can determine the velocity, or the momentum of the particle, we correct the expression in (A.12) by subtracting $\log h_p$ where $h_p$ is a small, but infinite interval:

$$H_{max}(momentum\ in\ 1D) = \frac{1}{2}\log(2\pi e m k_B T) - \log h_p \qquad (A.13)$$

Note again that if we choose the units of $h_p$ (of momentum as: $mass\ length/time$) the same as of $\sqrt{mk_BT}$, then the whole expression under the logarithm will be a pure number.

### A.3. Combining the SMI for the Location and Momentum of one Particle in 1D System

In the previous two sections, we derived the expressions for the locational and the momentum SMI of one particle in 1D system. We now combine the two results. Assuming that the location and the momentum (or velocity) of the particles are independent events we write

$$H_{max}(location\ and\ momentum) = H_{max}(location) + H_{max}(momentum)$$

$$= \log\left[\frac{L\sqrt{2\pi emk_BT}}{h_x h_p}\right] \tag{A.14}$$

Recall that $h_x$ and $h_p$ were chosen to eliminate the divergence of the SMI for a continuous random variables; location and momentum.

In (A.14) we assume that the location and the momentum of the particle are independent. However, quantum mechanics imposes restriction on the accuracy in determining both the location $x$ and the corresponding momentum $p_x$. In Equations (A.4) and (A.13) $h_x$ and $h_p$ were introduced because we did not care to determine the location and the momentum with an accuracy greater that $h_x$ and $h_p$, respectively. Now, we must acknowledge that nature imposes upon us a limit on the accuracy with which we can determine both the location and the corresponding momentum. Thus, in Equation (A.14), $h_x$ and $h_p$ cannot both be arbitrarily small, but their product must be of the order of Planck constant $h = 6.626 \times 10^{-34}\ J\ s$. Thus we set:

$$h_x h_p \approx h \tag{A.15}$$

And instead of (A.14), we write:

$$H_{max}(location\ and\ momentum) = \log\left[\frac{L\sqrt{2\pi emk_BT}}{h}\right] \tag{A.16}$$

### A.4. The SMI of a Particle in a Box of Volume V

We consider again one simple particle in a box of volume $V$. We assume that the location of the particle along the three axes $x$, $y$ and $z$ are independent. Therefore, we can write the SMI of the location of the particle in a cube of edges $L$, and volume $V$ as:

$$H(location\ in\ 3D) = 3H_{max}(location\ in\ 1D) \tag{A.17}$$

Similarly, for the momentum of the particle we assume that the momentum (or the velocity) along the three axes $x$, $y$ and $z$ are independent. Hence, we write:

$$H_{max}(momentum\ in\ 3D) = 3H_{max}(momentum\ in\ 1D) \tag{A.18}$$

We combine the SMI of the locations and momenta of one particle in a box of volume $V$, taking into account the uncertainty principle. The result is

$$H_{max}(location \ and \ momentum \ in \ 3D) = 3\log[\frac{L\sqrt{2\pi emk_BT}}{h}] \qquad (A.19)$$

### A.5. The SMI of Locations and Momenta of N indistinguishable Particles in a Box of Volume V

The next step is to proceed from one particle in a box to $N$ independent particles in a box of volume $V$. Giving the location $(x, y, z)$, and the momentum $(p_x, p_y, p_z)$ of one particle within the box, we say that we know the *microstate* of the particle. If there are $N$ particles in the box, and if their microstates are independent, we can write the SMI of $N$ such particles simply as $N$ times the SMI of one particle, i.e.,

$$\text{SMI(of } N \text{ independent particles)} = N \times \text{SMI(one particle)} \qquad (A.20)$$

This Equation would have been correct when the microstates of all the particles where independent. In reality, there are always correlations between the microstates of all the particles; one is due to *intermolecular interactions* between the particles, the second is due to the *indistinguishability* between the particles. We shall discuss these two sources of correlation separately.

*(i) correlation due to indistinguishability*

Recall that the microstate of a single particle includes the location and the momentum of that particle. Let us focus on the location of one particle in a box of volume $V$. We have written the locational SMI as:

$$H_{max}(location) = \log V \qquad (A.21)$$

Recall that this result was obtained for the continuous locational SMI. This result does not take into account the divergence of the limiting procedure. In order to explain the source of the correlation due to indistinguishability, suppose that we divide the volume $V$ into a very large number of small cells each of the volume $V/M$. We are not interested in the exact location of each particle, but only in which cell each particle is. The total number of cells is $M$, and we assume that the total number of particles is $N \ll M$. If each cell can contain at most one particle, then there are $M$ possibilities to put the first particle in one of the cells, and there are $M - 1$ possibilities to put the second particle in the remaining empty cells. Altogether, we have $M(M - 1)$ possible microstates, or configurations for two particles. The probability that one particle is found in cell $i$, and the second in a different cell $j$ is:

$$\Pr(i, j) = \frac{1}{M(M-1)} \qquad (A.22)$$

The probability that a particle is found in cell $i$ is:

$$\Pr(j) = \Pr(i) = \frac{1}{M} \qquad (A.23)$$

Therefore, we see that even in this simple example, there is correlation between the events "one particle in $i$" and one particle in $j$":

$$g(i, j) = \frac{\Pr(i,j)}{\Pr(i)\Pr(j)} = \frac{M^2}{M(M-1)} = \frac{1}{1 - \frac{1}{M}} \qquad (A.24)$$

Clearly, this correlation can be made as small as we wish, by taking $M \gg 1$ (or in general, $M \gg N$). There is another correlation which we cannot eliminate and is due to the indistinguishability of the particles.

Note that in counting the total number of configurations we have implicitly assumed that the two particles are labeled, say red and blue. In this case we count the two configurations in Figure 6a, as *different* configurations: "blue particle in cell $i$, and red particle in cell $j$," and "blue particle in cell $j$, and red particle in cell $i$."

Atoms and molecules are indistinguishable by nature; we cannot label them. Therefore, the two microstates (or configurations) in Figure 6b are indistinguishable. This means that the total number of configurations is not $M(M-1)$, but:

$$number\ of\ configurations = \frac{M(M-1)}{2} \to \frac{M^2}{2}, \text{for large } M \qquad (A.25)$$

For very large $M$ we have a correlation between the events "particle in $i$" and "particle in $j$":

$$g(i,j) = \frac{\Pr(i,j)}{\Pr(i)\Pr(j)} = \frac{M^2}{M^2/2} = 2 \qquad (A.26)$$

For $N$ particles distributed in $M$ cells, we have a correlation function (For $M \gg N$):

$$g(i_1, i_2, \ldots, i_n) = \frac{M^N}{M^N/N!} = N! \qquad (A.27)$$

This means that for $N$ indistinguishable particles we must divide the number of configurations $M^N$ by $N!$. Thus in general by removing the "labels" on the particles the number of configurations is *reduced* by $N!$. For two particles the two configurations shown in Figure 6a reduce to one shown in Figure 6b.

Now that we know that there are correlations between the events "one particle in $i_1$", "one particle in $i_2$" ... "one particle in $i_n$", we can define the *mutual information* corresponding to this correlation. We write this as:

$$I(1; 2; \ldots; N) = \ln N! \qquad (A.28)$$

The SMI for $N$ particles will be:

$$H(N\ particles) = \sum_{i=1}^{N} H(one\ particle) - \ln N! \qquad (A.29)$$

For the definition of the mutual information, see refs: [20, 24].

Using the SMI for the location and momentum of $N$ independent particles in (A.20) we can write the final result for the SMI of $N$ indistinguishable (but non-interacting) particles as:

$$H(N \text{ indistinguishable}) = N\log V \left(\frac{2\pi m e k_B T}{h^2}\right)^{3/2} - \log N! \qquad (A.30)$$

Using the Stirling approximation for $\log N!$ in the form (note again that we use the natural logarithm):

$$\log N! \approx N\log N - N \qquad (A.31)$$

We have the final result for the SMI of *N* indistinguishable particles in a box of volume *V*, and temperature *T*:

$$H(1,2,\ldots N) = N\log\left[\frac{V}{N}\left(\frac{2\pi m k_B T}{h^2}\right)^{3/2}\right] + \frac{5}{2}N \qquad (A.32)$$

By multiplying the SMI of *N* particles in a box of volume *V* at temperature *T*, by the factor $(k_B \log_e 2)$, one gets the *entropy*, the *thermodynamic entropy* of an ideal gas of simple particles. This equation was derived by Sackur and by Tetrode in 1912, by using the Boltzmann definition of entropy [45, 46].

One can convert this expression into the entropy function $S(E,V,N)$, by using the relationship between the total energy of the system, and the total kinetic energy of all the particles:

$$E = N\frac{m\langle v\rangle^2}{2} = \frac{3}{2}Nk_B T \qquad (A.33)$$

The explicit entropy function of an ideal gas is:

$$S(E,V,N) = Nk_B \ln\left[\frac{V}{N}\left(\frac{E}{N}\right)^{3/2}\right] + \frac{3}{2}k_B N\left[\frac{5}{3} + \ln\left(\frac{4\pi m}{3h^2}\right)\right] \qquad (A.34)$$

### (ii) Correlation Due to Intermolecular Interactions

In Equation (A.34) we got the entropy of a system of non-interacting simple particles (ideal gas). In any real system of particles, there are some interactions between the particles. Without getting into any details on the function $U(r)$, it is clear that there are two regions of distances $0 \leq r \lesssim \sigma$ and $0 \leq r \lesssim \infty$, where the slope of the function $U(r)$ is negative and positive, respectively. Negative slope correspond to repulsive forces between the pair of the particles when they are at a distance smaller than $\sigma$. This is the reason why $\sigma$ is sometimes referred to as the *effective diameter* of the particles. For larger distances, $r \gtrsim \sigma$ we observe attractive forces between the particles.

Intuitively, it is clear that interactions between the particles induce *correlations* between the locational probabilities of the two particles. For hard-spheres particles there is infinitely strong repulsive force between two particles when they approach to a distance of $r \leq \sigma$. Thus, if we know the location $\boldsymbol{R_1}$ of one particle, we can be sure that a second particle, at $\boldsymbol{R_2}$ is not in a sphere of diameter $\sigma$ around the point $\boldsymbol{R_1}$. This *repulsive* interaction may be said to introduce *negative correlation* between the locations of the two particles.

On the other hand, two argon atoms *attract* each other at distances $r \lesssim 4\text{Å}$. Therefore, if we know the location of one particle say, at $\boldsymbol{R_1}$, the probability of observing a second particle at $\boldsymbol{R_2}$ is *larger* than the probability of finding the particle at $\boldsymbol{R_2}$ in the absence of a particle at $\boldsymbol{R_1}$. In this case we get *positive correlation* between the locations of the two particles.

We can conclude that in both cases (attraction and repulsion) there are correlations between the particles. These correlations can be cast in the form of *mutual information* which reduces the SMI of a system of *N* simple particles in an ideal gas. The mathematical details of these correlations are discussed ref [20, 24].

Here, we show only the form of the mutual information at very low density. At this limit, we can assume that there are only *pair correlations*, and neglect all higher order correlations. The mutual information due to these correlations is:

$$I(due\ to\ correlations\ in\ pairs)$$

$$= \frac{N(N-1)}{2} \int p(\boldsymbol{R_1}, \boldsymbol{R_2}) \log g(\boldsymbol{R_1}, \boldsymbol{R_2}) d\boldsymbol{R_1} d\boldsymbol{R_2} \qquad (A.35)$$

where $\boldsymbol{g}(R_1, R_2)$ is defined by:

$$g(\boldsymbol{R_1}, \boldsymbol{R_2}) = \frac{p(\boldsymbol{R_1}, \boldsymbol{R_2})}{p(\boldsymbol{R_1}) p(\boldsymbol{R_2})} \qquad (A.36)$$

Note again that log g can be either positive or negative, but the average in (A.36) must be positive

## A.6 Conclusion

We summarize the main steps leading from the SMI to the entropy. We started with the SMI associated with the *locations* and *momenta* of the particles. We calculated the distribution of the locations and momenta that *maximizes* the SMI. We referred to this distribution as the *equilibrium distribution*. Let us denote this distribution of the locations and momenta of all the particles by $f_{eq}(\boldsymbol{R}, \boldsymbol{p})$.

Next, we use the equilibrium distribution to calculate the SMI of a system of *N* particles in a volume *V*, and at temperature *T*. This SMI is, up to a multiplicative constant ($k_B \ln 2$) identical with the *entropy* of an ideal gas at equilibrium. This is the reason we referred to the distribution which maximizes the SMI as the *equilibrium distribution*.

It should be noted that in the derivation of the entropy, we used the SMI twice; first, to calculate the distribution that maximize the SMI, then evaluating the maximum SMI corresponding to this distribution. The distinction between the concepts of SMI and entropy is essential. Referring to SMI (as many do) as entropy, inevitably leads to such an awkward statement: the maximal value of the entropy (meaning the SMI) is the entropy (meaning the thermodynamic entropy). The correct statement is that the SMI associated with locations and momenta is defined for any system; small or large, at equilibrium or far from equilibrium. This SMI, not the entropy, evolves into a maximum value when the system reaches equilibrium. At this state, the SMI becomes proportional to the entropy of the system.

Since the entropy is a special case of a SMI, it follows that whatever interpretation one accepts for the SMI, it will be automatically applied to the concept of entropy. The most important conclusion is that entropy is not a function of time. Entropy does not change with time, and entropy does not have a tendency to increase.

We said that the SMI may be defined for a system with any number of particles including the case $N = 1$. This is true for the SMI. When we talk about the entropy of a system we require that the system be very large. The reason is that only for such systems the entropy-formulation of the Second Law of thermodynamic is valid.

## Appendix B. The main assumption of Non-equilibrium Thermodynamics

Non-equilibrium thermodynamics is founded on the assumption of *local equilibrium* [51-54]. While it is true that this assumption leads to the entire theory of thermodynamics of non-equilibrium process, it is far from clear that the very assumption of local equilibrium can be justified.

Most textbooks on non-equilibrium thermodynamics starts with the reasonable assumption that in such system the intensive variables such as temperature $T$, pressure $P$, and chemical potential $\mu$ may be defined operationally in each small element of volume $dV$ of the system. Thus, one writes

$$T(\boldsymbol{R}, t), P(\boldsymbol{R}, t), \mu(\boldsymbol{R}, t) \qquad \text{(B.1)}$$

where $\boldsymbol{R}$ is the locational vector of a point in the system, and $t$ is the time. One can further assume that the density $\rho(\boldsymbol{R}, t)$ is defined locally at point $\boldsymbol{R}$ and integrate over the entire volume to obtain the total number of particles $N$

$$N = \int \rho(\boldsymbol{R}, t) d\boldsymbol{R} \qquad \text{(B.2)}$$

Similarly, one can define densities $\rho_R(\boldsymbol{R}, t)$ for each component of the system.

One also defines the local internal energy per unit of volume $u(\boldsymbol{R}, t)$. It is not clear however, how to integrate $u(\boldsymbol{R}, t)$ over the entire volume of the system to obtain the total internal energy of the system. While this may be done exactly for ideal gases, i.e. assuming that the total energy of the system is the sum of all the kinetic (as well as internal) energies of all the particles in the system, it is not clear how to do the same for systems having interacting particles. For suppose we divide the total volume of the system into $c$ small cells, Figure. 7, and assume that the internal energy in cell $i$ is $u(i, t)$. In which case the total internal energy of the system is written as

$$U = \sum_i u(i, t) \qquad \text{(B.2)}$$

And the corresponding integral is

$$U = \int u(\boldsymbol{R}, t) d\boldsymbol{R} \qquad \text{(B.4)}$$

Here, the integration is essentially the sum of the $u(\boldsymbol{R}, t)$ in small cells in the system, neglecting the interaction energies between the different cells. If there are interactions among the particles, then the internal energy of the system cannot be written as a sum of the form (B.3) or (B.4).

The most important and unjustified assumption is it is related to the local entropy $s(\boldsymbol{R}, t)$. One assumes that the entropy function, say $S(U, V, N)$ is the same function for the local entropy, i.e. $s$ is the same function as the local energy, volume, and number of particles of each element of volume.

Thus, the local entropy of the cell $i$ is presumed to be the same function of the local energy, volume, and number of particles, i.e. one writes

$$S = \sum_i s(i, t) = \int_V s(\boldsymbol{R}, t) d\boldsymbol{R} \qquad \text{(B.5)}$$

This assumption may be justified for ideal gas when the distribution of locations and velocities is meaningful for each element of volume in the system. To the best of the author's knowledge this assumption has never been justified for systems of interacting particles. The main problem in these definitions is that it does not take into account the correlations between different cells or between different elements of volume in the continuous case. These correlations depend on the number of particles in each cell which changes with time.

Once one makes such an assumption one can write the changes in the entropy of the system as

$$dS = d_e S + d_i S \qquad (B.6)$$

where $d_e S$ is the entropy change due to the heat exchange between the system and its surrounding, and $d_i S$ is the entropy produced in the system. For an isolated system $d_e S$, and all the entropy change is due to $d_i S$. The latter is further written as

$$\frac{d_i S}{dt} = \int \sigma d\boldsymbol{R} \qquad (B.7)$$

where $\sigma(\boldsymbol{R}, t)$ is the *local* entropy production

$$\sigma(\boldsymbol{R}, t) = \frac{d_i s}{dt} \geq 0 \qquad (B.8)$$

Thus, for an isolated system one has a local entropy production which is a function of time, and after integration one obtains also the total change of the entropy of the system as a function of time. Since the quantity $\sigma$ is defined in terms of the local entropy function $s(\boldsymbol{R}, t)$, and since $s(\boldsymbol{R}, t)$ is not a well-defined quantity, one should doubt the whole theory based on the assumption of local equilibrium.

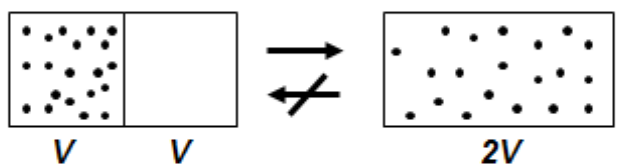

Figure 1. Expansion of an ideal gas from V to 2V.

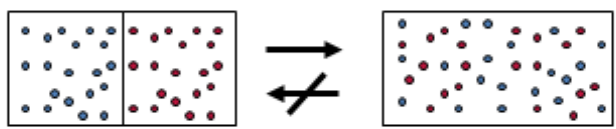

Figure 2. Mixing of two ideal gases

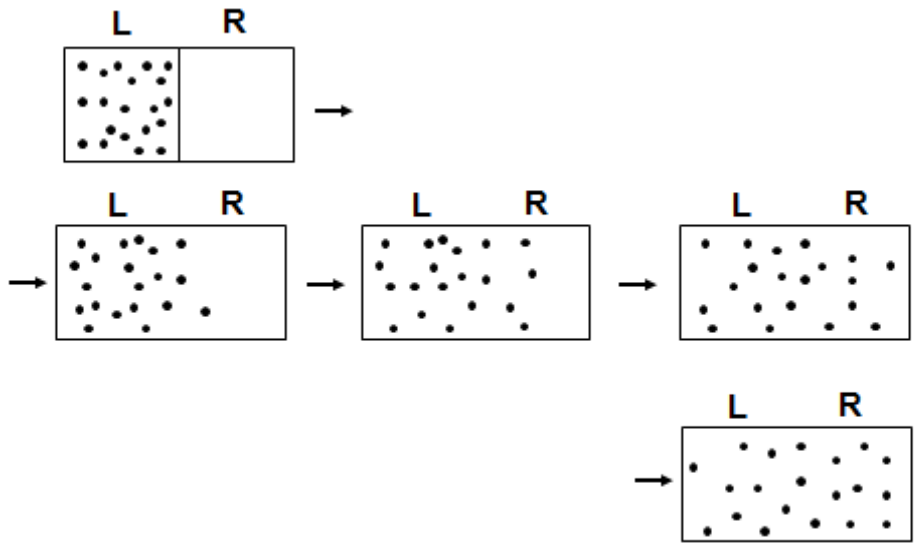

Figure 3. The initial, the final and some intermediate states
in the expansion process of Figure 1

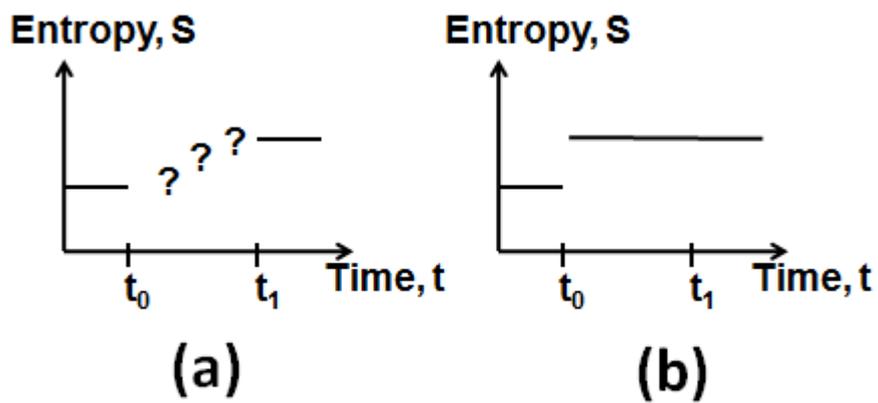

Figure 4 Two views of the entropy change after removal of the partition

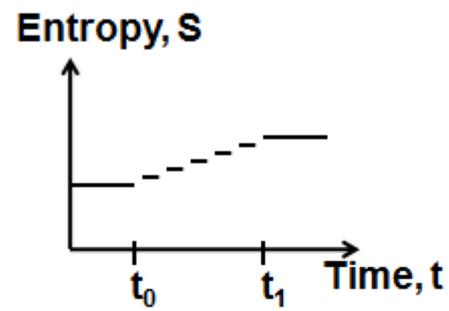

Figure 5. The gradual entropy change in a quasi-static process

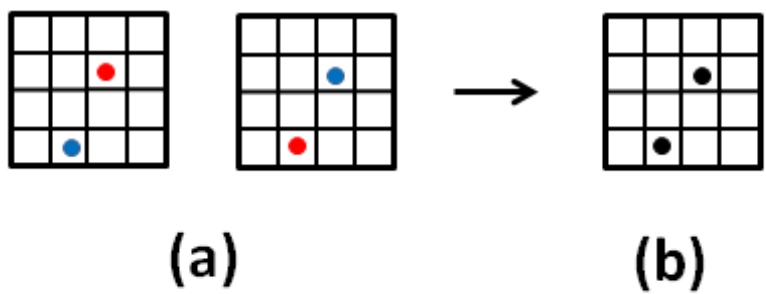

Figure 6. (a) Two distinguishable particles form two
different configurations, become one configuration
when the particles are indistinguishable (b)

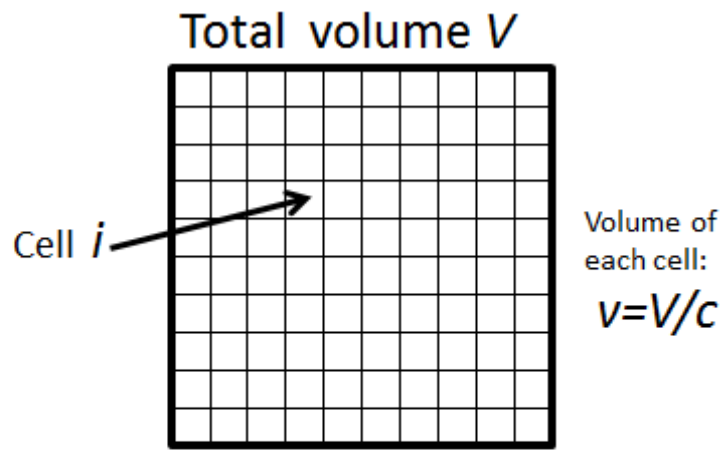

Figure 7. The whole macroscopic system of volume *V is* divided into c small cells each of volume *v*